\documentclass[runningheads]{llncs}

\usepackage[hidelinks]{hyperref}

\usepackage[draft]{minted}
\newenvironment{code}{\VerbatimEnvironment\begin{minted}{haskell}}{\end{minted}}
\newenvironment{spec}{\VerbatimEnvironment\begin{minted}{haskell}}{\end{minted}}
\newenvironment{js}{\VerbatimEnvironment\begin{minted}{javascript}}{\end{minted}}
\newcommand{\hs}{\mintinline{haskell}}

\usepackage[misc]{ifsym}

\usepackage{graphicx}

\usepackage{subcaption}

\title{PaSe: An Extensible and Inspectable DSL\\for Micro-Animations}

\long\def\ignore#1{}

\newcommand{\figscale}{0.68}

\newcommand{\dsl}{PaSe}

\begin{document}

\author{Ruben~P. Pieters $^{\textrm{\Letter}}$ \orcidID{0000-0003-0537-9403} \and
Tom Schrijvers \orcidID{0000-0001-8771-5559}}

\authorrunning{R. P. Pieters and T. Schrijvers}

\institute{KU Leuven, 3001 Leuven, Belgium
\\\email{\{ruben.pieters, tom.schrijvers\}@cs.kuleuven.be}}

\maketitle

\begin{abstract}
This paper presents \dsl{}, an extensible and inspectable DSL embedded in Haskell for expressing micro-animations. The philosophy of \dsl{} is to compose animations based on sequential and parallel composition of smaller animations. This differs from other animation libraries that focus more on sequential composition and have only limited forms of parallel composition. To provide similar flexibility as other animation libraries, \dsl{} features extensibility of operations and inspectability of animations.
We present the features of \dsl{} with a to-do list application, discuss the \dsl{} implementation, and argue that the callback style of extensibility is detrimental for correctly combining \dsl{} features. We contrast with the GreenSock Animation Platform, a professional-grade and widely used JavaScript animation library, to illustrate this point.
\end{abstract}

\section{Introduction}
\label{sec:intro}

Monads quickly became ubiquitous in functional programming because of their ability to structure effectful code in a pure functional codebase \cite{DBLP:conf/lfp/Wadler90}. However, monads have two major drawbacks. First, monads are not trivially extensible. A variety of techniques were developed to resolve this, including monad transformers \cite{DBLP:conf/popl/LiangHJ95}, free monads, and algebraic effects and handlers \cite{DBLP:conf/esop/PlotkinP09}. Second, monadic computations can only be inspected up to the next action. Techniques such as applicative functors \cite{DBLP:journals/jfp/McbrideP08}, arrows \cite{DBLP:journals/scp/Hughes00}, or selective applicative functors \cite{Mokhov:2019:SAF:3352468.3341694} increase the inspection capabilities by reducing the expressivity compared to monads.

This paper develops a domain specific language (DSL) embedded in Haskell for defining micro-animations, called \dsl{}\footnote{Pronounced \textit{pace}, the name is derived from \textit{\textbf{pa}rallel} and \textit{\textbf{se}quential}.}. \dsl{} employs the aforementioned techniques to support its key features: extensibility of operations and inspectability of animations while providing the freedom to express arbitrary animations.

Micro-animations are short animations displayed when users interact with an application, for example an animated transition between two screens. When used appropriately, they aid the user in understanding evolving states of the application \cite{DBLP:conf/infovis/BedersonB99,DBLP:conf/chi/Gonzalez96,DBLP:journals/tvcg/HeerR07}. Examples can be found in almost every software application: window managers shrink minimized windows and move them towards the taskbar, menus in mobile applications pop in gradually, browsers highlight newly selected tabs with an animation, etc.

This paper develops a DSL which combines various features expected of animation libraries, by building them on top of recent ideas found in functional programming. More concretely, our contributions are as follows:
\begin{itemize}
\item We develop \dsl{} which supports correct interactions for expressing arbitrary animations and inspectability. Animation libraries, such as the GreenSock Animation Platform (GSAP)\footnote{\url{https://greensock.com}}, typically use callbacks as a means of extensibility/expressivity which is detrimental to inspectability. We show an example where this results in unexpected behaviour, and how to correct it.
\item We extend and develop \dsl{} as an example of an \emph{extensible} DSL, in the sense that operations can be freely added. While there is a variety of literature on approaches to extensibility, there is not much literature devoted to use cases.
\item Animations in \dsl{} can be specified in an \emph{inspectable} manner, in the sense that information is obtained from a computation without running it. While inspectability has been studied for specific types of computations, such as free applicative functors \cite{DBLP:journals/corr/CapriottiK14}, it is a novel aspect to combine it with extensibility.
\item \dsl{} contains a primitive for parallel animation composition. This composition form can be arbitrarily nested and embedded anywhere within an animation and interacts correctly with the other features of \dsl{}: extensibility and inspectability. Parallel components in sequentially specified components is not new, see for example the \emph{par} element in UML sequence diagrams \cite{umlspec}, or the parallel statement in Ren'Py\footnote{\url{https://www.renpy.org/doc/html/atl.html\#parallel-statement}} or React Native Animations\footnote{\url{https://facebook.github.io/react-native/docs/animated\#parallel}}. But, those systems either do not provide a form of parallelism as general or interaction with other features such as inspectability is not present.
\end{itemize}

\section{Motivation}
\label{sec:motivation}

This section presents a simplified to-do list application to showcase the
functionality and features of \dsl{}.

\subsection{Running Example}

\begin{figure}[!htbp]
\centering
\includegraphics[width=\figscale\textwidth]{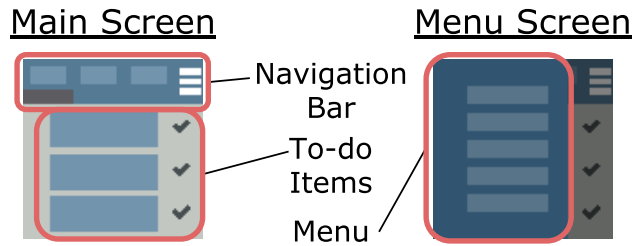}
\caption{Overview of the to-do list application.}
\label{fig:appOverview}
\end{figure}

Our application has two screens: a main screen and a menu screen. The main
screen contains a navigation bar and three items. An overview of the
application is given in Figure~\ref{fig:appOverview}.

\begin{figure}[!htbp]
\centering

\begin{subfigure}[h]{\textwidth}
\centering
\includegraphics[width=\figscale\textwidth]{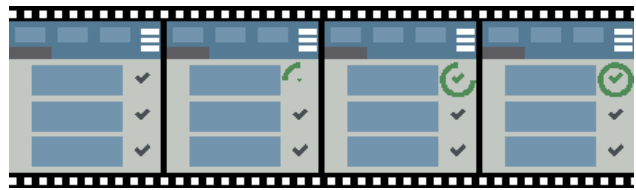}
\caption{\hs{markAsDone}: the checkmark changes shape and color.}
\label{fig:completeIconCheck}
\end{subfigure}

\begin{subfigure}[h]{\textwidth}
\centering
\includegraphics[width=\figscale\textwidth]{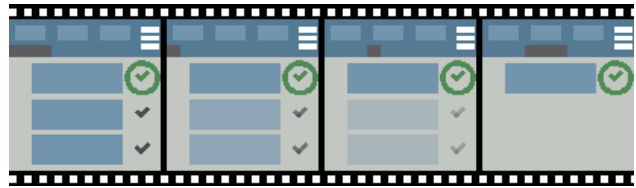}
\caption{\hs{onlyDone}: the to-do items fade out and the navbar underline changes.}
\label{fig:onlyDoneFig}
\end{subfigure}

\begin{subfigure}[h]{\textwidth}
\centering
\includegraphics[width=\figscale\textwidth]{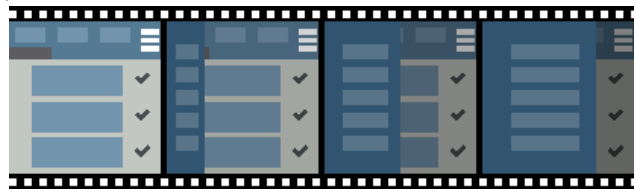}
\caption{\hs{menuIntro}: the menu appears while the background fades out.}
\label{fig:menuIntroFig}
\end{subfigure}

\caption{Micro-Animations in the to-do list application.}
\label{fig:animExamples}
\end{figure}

In this application, various user actions are accompanied with an animation:
\begin{itemize}
\item Each item can be marked as \emph{done} or \emph{not done} by clicking on it. The
checkmark icon changes shape and color to indicate the change in status. These
are the \hs{markAsDone}/\hs{markAsToDo} animations, of which \hs{markAsDone} is
shown in Figure~\ref{fig:completeIconCheck}.
\item Items can be filtered by their status by using the navigation bar
buttons. The first button shows all items, the second shows all done items, and
the third shows all to-do items. Both the navigation bar underline and the
to-do items itself change shape to indicate the change in selection. These are
the \hs{showAll}/\hs{onlyDone}/\hs{onlyToDo} animations, of which
\hs{onlyDone} is shown in Figure~\ref{fig:onlyDoneFig}.
\item The menu screen shows/hides itself after clicking the hamburger icon. The
menu expands inward from the left, to indicate the change in application state.
These are the \hs{menuIntro}/\hs{menuOutro} animations, of which \hs{menuIntro}
is shown in Figure~\ref{fig:menuIntroFig}.
\end{itemize}

\subsection{Composing Animations}

We express animations in \dsl{} by composing smaller animations into larger
ones. When creating an animation, we need come up with a suitable decomposition.
For example, the \hs{menuIntro} animation both introduces the menu screen and
fades out the background. Thus, it is composed of two basic animations
\hs{menuSlideIn} and \hs{appFadeOut} in parallel. The next sections explain how
to construct such basic and composed animations.

\subsection{Basic Animations}

A basic animation changes the property of an element over a period of time. To
specify a basic animation we need three elements: a lens specifying which
property to change, the duration of the animation, and lastly the target value
for this property. Since this operation creates an animation which changes a
property \emph{to} a target linearly, it is called \hs{linearTo}. The duration
is specified with the \hs{For} constructor while the target value is specified
with the \hs{To} constructor.

\paragraph{Note on Lenses} We use the lens notation \texttt{x . y . z} to
target the property \texttt{z} inside a nested structure \hs{{ x: { y: { z:
Property } } }}. This type of lenses was conceived by van Laarhoven
\cite{vlLenses}, and later packaged into
various Haskell libraries, such as
\texttt{lens}\footnote{\url{https://hackage.haskell.org/package/lens}}.

To implement the navigation bar underline animation, we reduce the underline
width of the first button for 0.25 seconds and increase the underline width of
the second button for 0.25 seconds. These animations are expressed in
respectively \hs{line1Out} and \hs{line2In} below, and shown visually in
Figure~\ref{fig:basic}.

\begin{spec}
line1Out = linearTo (navbar . underline1 . width) (For 0.25) (To 0)
line2In = linearTo (navbar . underline2 . width) (For 0.25) (To 28)
\end{spec}

Other examples are the \hs{menuSlideIn} and \hs{appFadeOut} animations. For the
former, we increase the width of the menu over a duration of 0.5 seconds, and
for the latter we increase the alpha value of the obscuring box over a duration
of 0.5 seconds. These animations are shown visually in Figure~\ref{fig:basic}.

\begin{spec}
menuSlideIn = linearTo (menu . width) (For 0.5) (To 75)
appFadeOut = linearTo (obscuringBox . alpha) (For 0.5) (To 0.65)
\end{spec}

\begin{figure}[!htbp]
\centering

\begin{subfigure}[h]{\textwidth}
\centering
\includegraphics[width=\figscale\textwidth]{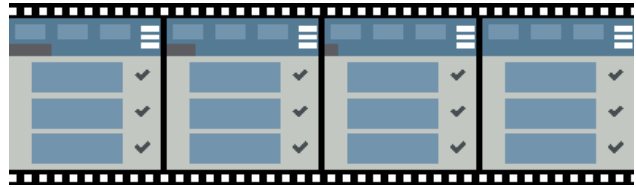}
\caption{The \hs{line1Out} animation.}
\label{fig:basic1_1}
\end{subfigure}

\begin{subfigure}[h]{\textwidth}
\centering
\includegraphics[width=\figscale\textwidth]{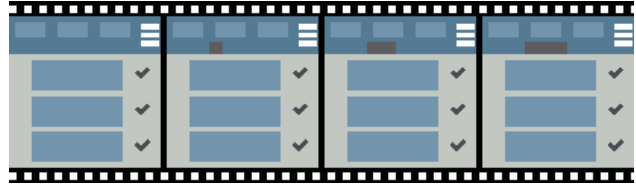}
\caption{The \hs{line2In} animation.}
\label{fig:basic1_2}
\end{subfigure}

\begin{subfigure}[h]{\textwidth}
\centering
\includegraphics[width=\figscale\textwidth]{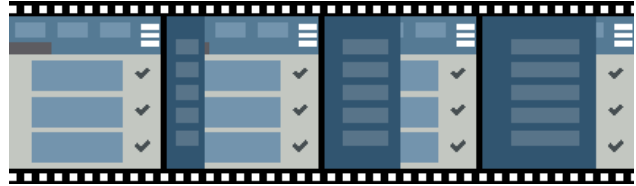}
\caption{The \hs{menuSlideIn} animation.}
\label{fig:basic2_1}
\end{subfigure}

\begin{subfigure}[h]{\textwidth}
\centering
\includegraphics[width=\figscale\textwidth]{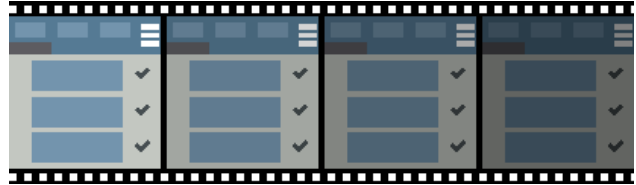}
\caption{The \hs{appFadeOut} animation.}
\label{fig:basic2_2}
\end{subfigure}

\caption{Basic \hs{linearTo} animations.}
\label{fig:basic}
\end{figure}

\subsection{Composed Animations}

A composed animation combines several other animations into one new animation. We can do this either in \emph{sequence} or in \emph{parallel}.

To obtain the \hs{selectBtn2} animation, we combine both the \hs{line1Out} and \hs{line2In} animations with the \hs{sequential} combinator. This constructs a new animation which first plays the \hs{line1Out} animation, and once it is finished plays the \hs{line2In} animation.

\begin{spec}
selectBtn2Anim = line1Out `sequential` line2In
\end{spec}

To obtain the \hs{menuIntro} animation, we combine both the \hs{menuSlideIn} and \hs{appFadeOut} animations with the \hs{parallel} combinator. This constructs a new animation which plays both the \hs{menuSlideIn} and \hs{appFadeOut} animations at the same time.

\begin{spec}
menuIntro = menuSlideIn `parallel` appFadeOut
\end{spec}

\begin{figure}[!htbp]
\centering

\begin{subfigure}[h]{\textwidth}
\centering
\includegraphics[width=\figscale\textwidth]{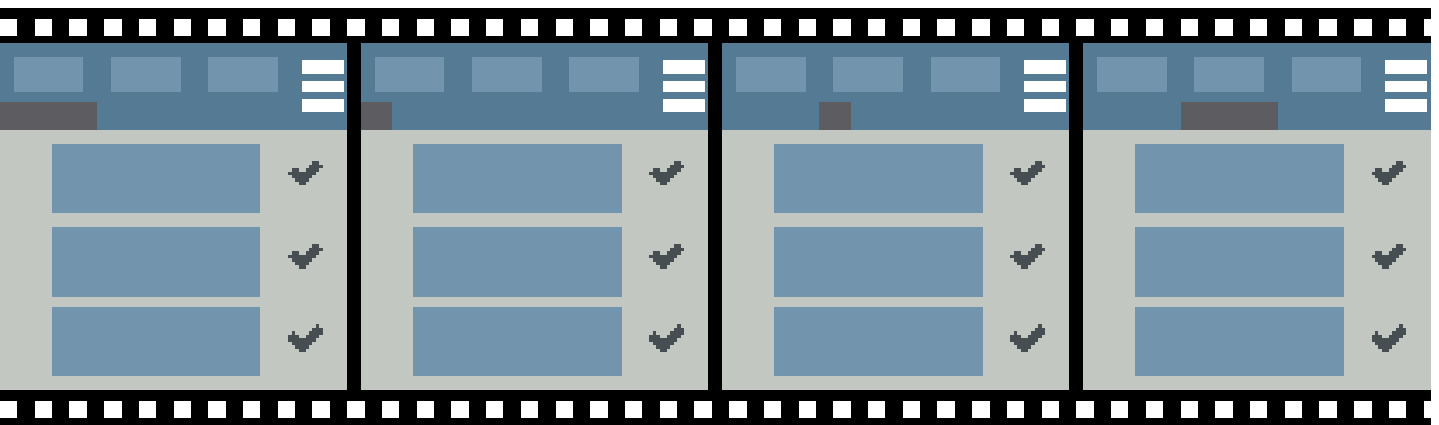}
\caption{The \hs{selectBtn2} animation.}
\label{fig:composed1}
\end{subfigure}

\begin{subfigure}[h]{\textwidth}
\centering
\includegraphics[width=\figscale\textwidth]{pictures/menuIntroFig}
\caption{The \hs{menuIntro} animation.}
\label{fig:composed2}
\end{subfigure}

\caption{All of the defined composed animations.}
\label{fig:composed}
\end{figure}

Both of these animations are shown visually in Figure~\ref{fig:composed}.

\section{Extensibility, Inspectability and Expressiveness}
\label{sec:features}

The features seen in Section~\ref{sec:motivation} form the basis of \dsl{}.
However, animation libraries such as GSAP provide a much more extensive list of
features.

To add support for similar features, we design \dsl{} with extensibility and
inspectability in mind. This means that \dsl{} can be extended with custom
operations and information can be derived from inspecting specified animations.
To support arbitrary expressiveness in combination with those
features, we also emphasize the possibility to extend \dsl{} with custom
combinators.

\subsection{Extensibility}
\label{sec:customop}

The \hs{linearTo} operation and the \hs{sequential} and \hs{parallel} combinators form the basis for expressing a variety of animations. However, there are different situations which require different primitives to express our desired animation. For example, GSAP provides a primitive to morph one shape into another.

An example in the to-do list app is the \hs{checkIcon} animation, part of the \hs{markAsDone} animation, where we want to set the color of the checkmark to a new value. We define a custom \hs{set} operation and embed it inside a \dsl{} animation. In this animation we use Haskell's \hs{do}-notation to specify sequential animations.

\begin{spec}
checkIcon = do ...; set (checkmark . color) green; ...
\end{spec}


\subsection{Inspectability}

\dsl{} is inspectable, meaning that we can derive properties of expressed computations by \emph{inspecting them rather than running them}. For example, if we want to know the duration of \hs{menuIntro} without actually running it and keeping track of the time. We can do this by using a predefined \hs{duration} function, which calculates the duration by inspecting the animation. This gives a duration of \hs{0.5} seconds, which is indeed the duration of two \hs{0.5} second animations in parallel.

\begin{spec}
menuIntroDuration = duration menuIntro -- = 0.5
\end{spec}

Of course, it is not possible to inspect every animation. In the following situation we have a custom operation \hs{get}, the dual of \hs{set} in the previous section, returning a \hs{Float}. If the result of this value is used as the duration parameter of an animation, then we cannot know upfront how long this animation will last. Requesting to calculate the duration of such an animation results in a type error.

\begin{spec}
complicatedAnim = do v <- get; linearTo lens (For v) (To 10)
complicatedAnimDuration = duration complicatedAnim -- type error
\end{spec}

Calculating a duration is mostly a stepping stone towards supporting other interesting features. One such example is sequentially composing animations with a relative offset. For example, to compose a first animation \hs{anim1} with a second animation \hs{anim2} which starts 0.5 seconds before the end of \hs{anim1}.

\begin{spec}
relSeqAnim = relSequential anim1 anim2 (-0.5)
\end{spec}

\subsection{Expressiveness}
\label{sec:customcomb}

Similarly to providing custom operations, \dsl{} also supports custom
combinators. For typical monadic DSLs this is not needed as \hs{>>=} and \hs{return} 
cover all use cases.
However, since \dsl{} has the additional requirement of being inspectable, the
\hs{>>=} combinator ends up being a liability because it only provides a very
limited amount of inspectability.

In the \hs{onlyDone} animation, we show all \emph{done} items, while
hiding all \emph{to-do} items. This could be implemented by first
showing all items with the \hs{showAll} animation, since an item might have
been hidden by a previous action, and then hiding all to-do items with the
\hs{hideToDo} animation. The definition for this is given below, while the
implementation of \hs{showAll} and \hs{hideToDo} is omitted.

\begin{spec}
onlyDoneNaive = do showAll; hideToDo
\end{spec}

However, this animation is a bit naive since it executes the \hs{showAll}
animation regardless of whether there are any hidden done items that actually
need to appear. Instead, the intended animation is to only make done items
appear when any were hidden. This is done by first checking whether there are
any done items and based on that we play the naive version of \hs{onlyDone},
otherwise we just hide the not done items.

\begin{spec}
onlyDone = do
  cond <- doneItemsGt0    -- check if more than 0 'done' items
  if cond then onlyDoneNaive else hideToDo
\end{spec}

However, if we also want to inspect this computation, this formulation
is problematic since it is specified using a construction which is too general;
we explain this in more detail in Section~\ref{sec:interaction}. Instead, we
can define a custom combinator \hs{ifThenElse} to express \hs{onlyDone}
in an inspectable manner.

\begin{spec}
onlyDone = ifThenElse doneItemsGt0 onlyDoneNaive hideToDo
\end{spec}

For this new combinator, we can define custom ways to inspect it. Since each
branch might have a different duration, we do not choose to extract the
duration but rather the \emph{maximum} duration of the animation.

\begin{spec}
onlyDoneMaxDuration = maxDuration onlyDone -- = 1
\end{spec}

Sections~\ref{sec:motivation} and \ref{sec:features} gave a look and feel of the
features of \dsl{}. In the following sections, we delve deeper into the
internals of the implementation.

\section{Implementation of \dsl{}}
\label{sec:detail}

This section implements the earlier introduced operations and redefines the animations to show the resulting type signature. We
develop \dsl{} in the style of the \texttt{mtl}
library\footnote{\url{http://hackage.haskell.org/package/mtl}} which implements
monadic effects using typeclasses \cite{DBLP:conf/afp/Jones95}.  This style is
also called the finally tagless approach \cite{DBLP:journals/jfp/CaretteKS09}.
However, because the \dsl{} classes are not subclasses of \hs{Monad} they leave
room for inspectability.

\subsection{Specifying Basic Animations}

The \texttt{mtl} library uses typeclasses to declare the basic operations of an
effect. Similarly, we specify the \hs{linearTo} operation using
the \hs{LinearTo} typeclass.

\begin{code}
class LinearTo obj f where
  linearTo :: Traversal' obj Float -> Duration -> Target -> f ()
\end{code}

The traditional mtl-style would add a \hs{Monad f} superclass constraint. Yet,
because it hinders inspectability, we do not require it everywhere. Instead,
while we can still choose to add an additional \hs{Monad f} constraint in selected
cases, we leave the freedom to opt for an alternative like \hs{Applicative f} in
other cases.

The \hs{linearTo} function of the \hs{LinearTo} typeclass specifies basic
animations like \hs{line1Out}, \hs{line2In},
\hs{menuSlideIn}, and \hs{appFadeOut} animations from
Section~\ref{sec:motivation}. 
As an example, we redefine \hs{line1Out} with its type signature; the others 
are similar. 

\begin{code}
line1Out :: (LinearTo Application f) => f ()
line1Out = linearTo (navbar . underline1 . width) (For 0.25) (To 0)
\end{code}

\subsection{Specifying Composed Animations}

Section~\ref{sec:motivation} used the combinators \hs{sequential} and
\hs{parallel} for composing animations. Here, we provide these as well as more
general combinators.


\subsubsection{Sequential Composition}

We reuse the \hs{Functor}-\hs{Applicative}-\hs{Monad} hierarchy for
sequencing animations. In particular, the \hs{sequential} function is a
specialization of the \hs{liftA2}
function\footnote{\url{https://hackage.haskell.org/package/base-4.12.0.0/docs/Control-Applicative.html\#v:liftA2}}.

\begin{code}
sequential :: (Applicative f) => f () -> f () -> f ()
sequential f1 f2 = liftA2 (\_ _ -> ()) f1 f2
\end{code}

Hence, the type signature for \hs{selectBtn2Anim} contains an 
\hs{Applicative f} constraint in addition to the \hs{LinearTo Application f}
constraint.

\begin{code}
selectBtn2Anim :: (LinearTo Application f, Applicative f) => f ()
selectBtn2Anim = line1Out `sequential` line2In
\end{code}

\subsubsection{Parallel Composition}

The \hs{parallel} function does not correspond to a member function in an
existing Haskell typeclass. At first we might consider the \hs{<|>} combinator
in the \hs{Alternative} class, but there is no sensible corresponding
\hs{empty} animation; also the suggested 
laws\footnote{\url{https://en.wikibooks.org/wiki/Haskell/Alternative_and_MonadPlus}}
do not make sense for animations.

Instead, we create our own \hs{Parallel} typeclass below. Its function
\hs{liftP2} has the same signature as \hs{liftA2}, but denotes
parallel composition instead of sequential composition. The
\hs{parallel} function is a specialization of \hs{liftP2}.

\begin{code}
class Parallel f where
  liftP2 :: (a -> b -> c) -> f a -> f b -> f c

parallel :: (Parallel f) => f () -> f () -> f ()
parallel f1 f2 = liftP2 (\_ _ -> ()) f1 f2
\end{code}
 and
Now we can give a type signature for \hs{menuIntro}.
\begin{code}
menuIntro :: (LinearTo Application f, Parallel f) => f ()
menuIntro = menuSlideIn `parallel` appFadeOut
\end{code}

\subsection{Running Animations}

Now we create a new \hs{Animation} datatype that instantiates the above
typeclasses to interpret \dsl{} programs as actual animations. We briefly
summarize this implementation here and refer for more details to
our prototype
implementation.\footnote{\url{https://github.com/rubenpieters/anim_eff_dsl/tree/master/code}}

The function type
\hs{obj}~\hs{->}~\hs{Float}~\hs{->}~\hs{m}~\hs{(}\hs{obj,}~\hs{(Either}~\hs{(Animation}~\hs{obj}~\hs{a)}~\hs{a))}
models an animation.
It takes as inputs the the current application state and the
time elapsed since the previous frame. As output it produces a new
application state for the next frame together with either the remainder of the animation or, if there is no remainder, the result of
the animation. Note that the output is wrapped in a type
constructor \hs{m} to embed custom effects. 
One more detail is missing in the output: the remaining unused time. We need
this when, for example, there is more time between frames than the animation
uses.
Then, the remaining time
can be used to run part of the next animation. All together, we get
the following type definition.

\begin{code}
newtype Animation obj m a = Animation { runAnimation ::
    obj ->                           -- previous state
    Float ->                         -- time delta
    m ( obj                          -- next state
      , Either (Animation obj m a) a -- remainder / result
      , Maybe Float )}               -- remaining time delta
\end{code}

\subsubsection{LinearTo Instance}

The \hs{linearTo} implementation of \hs{Animation} does three things: construct
the new object state, calculate the remainder of the animation, and calculate
the unused time. The difference between the \hs{linearTo} duration and the
frame time determines whether there is a remaining \hs{linearTo} animation or
remaining time.

\paragraph{Examples}

Let us illustrate the behaviour on a simple application
state that is a tuple, \hs{type State = (Float, Float)}, of an {x}-value and a {y}-value.
The \hs{right} animation transforms the \hs{x}-value to 50 over a time of 1
second.

\begin{code}
right :: (LinearTo State f) => f ()
right = linearTo x (For 1) (To 50)
\end{code}

We run the animation for 0.5 seconds by applying the \hs{runAnimation} function
on \hs{right}, together with the initial state (\hs{s0 =
(0,0)}) and the duration \hs{0.5}. We also instantiate the \hs{m} type constructor \hs{m} inside
the \hs{Animation} data type with \hs{Identity} as no additional effects are
needed; this means that the result can be unwrapped with
\hs{runIdentity}.

\begin{code}
(s1, remAn1, remDel1) = runIdentity (runAnimation right s0 0.5)
-- s1 = (25.0, 0.0) | remAn1 = Left anim2 | remDel1 = Nothing
\end{code}

Running \hs{right} for 0.5 seconds uses all available time and yields the new state
\hs{(25, 0)}. The remainder of
the animation is the \hs{right} animation with its duration reduced by \hs{0.5}, or essentially \hs{linearTo x (For 0.5) (To 50)}. Let us run this
remainder for 1 second.

\begin{code}
(s2, remAn2, remDel2) = runIdentity (runAnimation anim2 s1 1)
-- s2 = (50.0, 0.0) | remAn2 = Right () | remDel2 = Just 0.5
\end{code}

Now the animation finishes in state \hs{(50, 0)} with result \hs{()} and remaining time \hs{0.5}.

\subsubsection{Monad Instance}

For sequential animations we provide a \hs{Monad} instance. Its \hs{return} embeds the
result \hs{a} inside the \hs{Animation} data type. The essence of the
\hs{f}~\hs{>>=}~\hs{k} case is straightforward: first, run the animation
\hs{f}, then pass its result to the continuation \hs{k} and run that animation.
Of course, running \hs{f} does not necessarily give us a result \hs{a}. Running
an animation gives us either a result or an animation remainder; we also have
to take into account the potential remaining time. When we have an
animation remainder instead of a result, we need to repackage this remaining
animation as \hs{remainder}~\hs{>>=}~\hs{k} instead of using \hs{k}~\hs{a}.
Additionally, if there is no remaining time,  we do not continue
running the continuation and instead return the current state and animation
remainder.

\paragraph{Examples}

Let us define an additional animation \hs{up} which transforms the \hs{y}-value to 50 over a duration of 1 second. Additionally, we define an animation \hs{rightThenUp} which composes the \hs{right} and \hs{up} animations in sequence.

\begin{code}
up :: (LinearTo State f) => f ()
up = linearTo y (For 1) (To 50)

rightThenUp :: (LinearTo State f, Applicative f) => f ()
rightThenUp = right `sequential` up
\end{code}

Running the \hs{rightThenUp} animation for 0.5 seconds gives a similar result
to running \hs{right} for 0.5 seconds. We obtain the new state \hs{(25,
0)}, an animation remainder \hs{anim2} and there is no remaining time.
Now the animation remainder is the rest of the \hs{rightThenUp} animation,
which is half of the \hs{right} animation and the full \hs{up} animation. So,
when we run this animation remainder for 1 second, it will run the second half
of \hs{right} and the first half of \hs{up}.  This results in the state
\hs{(50, 25)}, the animation remainder \hs{anim3} and no remaining delta time.
This animation remainder is of course the second half of the \hs{up} animation.
If we continue to run that remainder, for example for 1 second, then we get the
final state \hs{(50, 50)} and the animation result \hs{()}.

\subsubsection{Parallel Instance}

For parallel animations the \hs{liftP2} implementation runs
the animations \hs{f1} and \hs{f2} both on the starting
state. We distinguish different cases depending on whether \hs{f1} and
\hs{f2} finish with a result or an animation remainder, and the remaining
time. We check which of the animations have finished and repackage them
either into a result or a new remainder, using the result combination function
where appropriate. When the longest of the two parallel animations is finished
while not fully using the remaining delta time, we continue running the remainder
of the animation.

\paragraph{Examples}

Let us run the animations \hs{right} and \hs{up} in parallel, which means
that both the \hs{x} and \hs{y}-value will increase simultaneously.

\begin{code}
rightAndUp :: (LinearTo State f, Parallel f) => f ()
rightAndUp = right `parallel` up
\end{code}

The result of running this animation for 0.5 seconds gives the state \hs{(25,
25)} and no remaining time. If we continue the animation
remainder we get the state \hs{(50, 50)} and 0.5 seconds of remaining 
time.

\subsection{Inspecting Animations}

Inspecting an animation is done by interpreting \dsl{} to a different data
type, which provides instances for the typeclasses with different semantics.
In particular we show how to inspect the duration of an animation
by means of the \hs{Const} functor.

\begin{spec}
newtype Const a b = Const { getConst :: a }
\end{spec}

\subsubsection{Inspecting LinearTo}

To inspect animations we provide an instance for the operations which
interprets them into \hs{Const Duration}. In the case of the \hs{linearTo}
operation, we simply embed the duration within a \hs{Const} wrapper.

\begin{code}
instance LinearTo obj (Const Duration) where
  linearTo _ duration _ = Const duration
\end{code}

\subsubsection{Inspecting Applicative}

While the \hs{Const} data type does not provide a \hs{Monad} instance, it does
provide an \hs{Applicative} instance. This means that it is not possible to
inspect animations with a \hs{Monad} constraint, but it is possible for
animations with an \hs{Applicative} constraint. The \hs{Const} data type is not
the culprit here, but rather the \hs{>>=} method of the \hs{Monad} class which
contains a continuation function \hs{a}~\hs{->}~\hs{m}~\hs{b}, which is the
limiting factor of inspection.

\subsubsection{Inspecting Parallel}

The duration of two parallel animations is the maximum of their durations. 
The \hs{Par} instance implementation for \hs{Const Duration} captures
this. 

\begin{code}
instance Par (Const Duration) where
  liftP2 _ (Const x1) (Const x2) = Const (max x1 x2)
\end{code}

\paragraph{Examples}

The duration function is a specialization of the unwrapper function of the
\hs{Const} data type, namely \hs{getConst}. We can feed our previously defined
animations \hs{selectBtn2Anim} and \hs{menuIntro} from
Section~\ref{sec:motivation} to this function and obtain the duration of the
animations as a result.

\begin{code}
duration :: Const Duration a -> Duration
duration = getConst

selectBtn2AnimDuration :: Duration
selectBtn2AnimDuration = duration selectBtn2Anim -- = For 1.0

menuIntroDuration :: Duration
menuIntroDuration = duration menuIntro -- = For 0.5
\end{code}

When we try to retrieve the duration of an animation with a monad constraint,
we receive an error from the compiler: it cannot find a \hs{Monad} instance for
\hs{Const Duration}.

\begin{spec}
complicatedAnimDuration :: Duration
complicatedAnimDuration = duration complicatedAnim
-- No instance for (Monad (Const Duration))
\end{spec}

\subsection{Adding a Custom Operation}

Adding a custom operation is done as in any other mtl-style approach: by defining a new class for this operation. For example, if we want to add a \hs{set} operation, then we create a corresponding \hs{Set} class.

\begin{code}
class Set obj f where set :: Lens' obj a -> a -> f ()
\end{code}

Now, an animation that uses the \hs{set} operation will incur a \hs{Set} constraint.

\begin{code}
checkIcon :: (Set CompleteIcon f, ...) => f ()
checkIcon = do ...; set (checkmark . color) green; ...
\end{code}

To inspect or run such an animation, we also need to provide instances for the \hs{Animation} and \hs{Const} data types. In the \hs{Animation} instance, we alter the previous state by setting the value targeted by the \hs{lens} to \hs{a}. The duration of a \hs{set} animation is 0, which is what is returned in the \hs{Duration} instance.

\begin{code}
instance (Applicative m) => Set obj (Animation obj m) where
  set lens a = Animation $ \obj t -> let
    newObj = Lens.set lens a obj
    in pure (newObj, Right (), Just t)

instance Set obj (Const Duration) where
  set _ _ = Const (For 0)
\end{code}

This section gave an overview of the features provided by \dsl{}, however combining inspectability while allowing freedom of expressivity is not always straightforward. Therefore, we look at an example of the interaction of these two features in the next section.

\section{Interaction Between Inspectability and Expressivity}
\label{sec:interaction}

Haskell DSLs are typically monadic because the \hs{>>=} combinator provides
great expressive power. Yet, this power also hinders inspectability. This section shows how to balance 
expressiveness and inspectability with a custom combinator.

Let us revisit the \hs{onlyDone} animation from Section~\ref{sec:customcomb}. The following definition imposes a \hs{Monad} constraint on \hs{f}, making the animation non-inspectable.

\begin{spec}
onlyDone :: (LinearTo Application f, Get Application f,
  Set Application f, Monad f, Parallel f) => f ()
onlyDone = do
  cond <- doneItemsGt0
  if cond then onlyDoneNaive else hideNotDone
\end{spec}

However, it does seem like we should be able to extract some duration related
information from it. For example, the maximum duration should be the largest
duration of the two branches.

To express this idea in \dsl{} we introduce an explicit combinator to replace
the particular use of \hs{>>=}, namely an \hs{if-then-else} construction. 

\begin{code}
class IfThenElse f where
  ifThenElse :: f Bool -> f a -> f a -> f a
\end{code}

This is similar to the \hs{handle} combinator from the \hs{DynamicIdiom} class
\cite{DBLP:phd/ethos/Yallop10} and the \hs{ifS} combinator from the
\hs{Selective} class \cite{Mokhov:2019:SAF:3352468.3341694}.

Now we can reformulate \hs{onlyDone} in terms of this \hs{ifThenElse} combinator.

\begin{code}
onlyDone :: (LinearTo Application f, Get Application f,
  Set Application f, Applicative f, Parallel f, IfThenElse f)
  => f ()
onlyDone = ifThenElse doneItemsGt0 onlyDoneNaive hideNotDone
\end{code}

Not much changes for \hs{Animation}, which implements \hs{ifThenElse} in terms of
the \hs{>>=} combinator. 

\begin{code}
instance (Monad f) => IfThenElse (Animation obj f) where
  ifThenElse fBool thenBranch elseBranch = do
    bool <- fBool
    if bool then thenBranch else elseBranch
\end{code}

What is new is that we can now also retrieve the maximum duration, using a new type
\hs{MaxDuration} to signify that we are not simply calculating the duration of
the animation. The instance for \hs{IfThenElse} retrieves the
durations of the \texttt{then} and \texttt{else} branches and adds the greater
value to the duration of the preceding animation inside the condition.

\begin{code}
instance IfThenElse (Const MaxDuration) where
  ifThenElse (Const (MaxDur durCond)) (Const (MaxDur durThen))
             (Const (MaxDur durElse)) =
    Const (MaxDur (durCond + max durThen durElse))
\end{code}

This allows us to retrieve the maximum duration of the \hs{onlyDone} animation.

\begin{spec}
onlyDoneMaxDuration :: MaxDuration
onlyDoneMaxDuration = maxDuration onlyDone -- = MaxDur 1.0
\end{spec}

\section{Interaction Between Callbacks and Inspectability}
\label{sec:evaluation}

Many JavaScript animation libraries\footnote{Examples:
\url{https://greensock.com}, \url{https://animejs.com}, and
\url{https://popmotion.io}.} exist, most of which allow the user to add custom
behavior (which the library has not foreseen) through callbacks. A good example
is the GreenSock Animation Platform (GSAP), a widely recommended and mature
JavaScript animation library with a variety of features.

\subsection{Working with GSAP}

GSAP provides primitives similar to the \hs{linearTo} and
\hs{sequential} operations in \dsl{}. While it does not have a primitive for
\hs{parallel} composition, we can define it in terms of other GSAP features.

\texttt{TweenMax} objects are the GSAP counterpart of the
\hs{linearTo} operation. Their arguments are similar: the object to change, the
duration, and the target value for the
property. For example, animation \texttt{right} moves
\texttt{box1} to the right:

\begin{js}
const right = new TweenMax("#box1", 1, { x: 50 });
\end{js}

A sequential animation is created by creating a timeline and \texttt{add}ing
animations to it.  Below, we create the animation \texttt{rightThenDown} which
moves \texttt{box1} to the right and then moves it down.

\begin{js}
const rightThenDown = new TimelineMax({ paused: true })
  .add(new TweenMax("#box1", 1, { x: 50 }))
  .add(new TweenMax("#box1", 1, { y: 50 }));
\end{js}

The \texttt{add} method takes the position on the timeline as an optional
paramter. Hence, if we position both animations at point \texttt{0} on the timeline,
they will run in parallel. For example, the \texttt{both} animation below 
moves both \texttt{box1} and \texttt{box2} in parallel to an \texttt{x}-value
of 50.

\begin{js}
const both = new TimelineMax({ paused: true })
  .add(new TweenMax("#box1", 1, { x: 50 }), 0)
  .add(new TweenMax("#box2", 1, { x: 50 }), 0);
\end{js}

Since we have specified the option \texttt{paused} with \texttt{true}, we must
call the \texttt{play} method to run an animation, as in
\texttt{both.play()}. Using this behaviour, and the ability to nest timelines,
we can provide the \dsl{} primitives.

\begin{js}
function sequential(tl1, tl2) {
  return new TimelineMax({ paused: true })
    .add(tl1.play())
    .add(tl2.play()); }

function parallel(tl1, tl2) {
  return new TimelineMax({ paused: true })
    .add(tl1.play(), 0)
    .add(tl2.play(), 0); }
\end{js}

\subsection{Callbacks and Inspectability}

GSAP also provides features related to inspectability. For example, we can use the
\texttt{totalDuration} method to return the total duration of an animation.
Animations created with the previously defined \texttt{sequential} and
\texttt{parallel} functions correctly give their total duration when queried.
For example, we define an animation which moves both boxes in parallel to
\texttt{x = 50} and then back to \texttt{x = 0} in parallel. The duration is
correctly returned as 2.

\begin{js}
const bothTo50 = parallel(
  new TweenMax("#box1", 1, { x: 50 }),
  new TweenMax("#box2", 1, { x: 50 }));

const bothTo0 = parallel(
  new TweenMax("#box1", 1, { x: 0 }),
  new TweenMax("#box2", 1, { x: 0 }));

const bothAnimation = sequential(bothTo50, bothTo0);
const bothAnimDuration = bothAnimation.totalDuration(); // = 2
\end{js}

However, if we want to provide animations similar to \hs{onlyDone}, which contains an \hs{if-then-else}, then the duration returned is not what we expect. The \texttt{add} method is overloaded to take a callback as parameter, which we used in the previous section since \texttt{tl.play()} is shorthand for the callback \texttt{() => tl.play()}. Using the callback parameter we can embed arbitrary effects and control flow. For example, we can create a conditional animation \texttt{condAnim}, for which a duration of 1 is returned.This is different from the expected total duration of the animation, which is 2. (Of course, in general the duration of the animations in both branches could differ, which is what makes it difficult to provide a procedure for calculating the duration of an animation in this form.)

\begin{js}
const condAnim = new TimelineMax({ paused: true })
  .add(bothTo50.play())
  .add(() => { if (cond) { bothTo0.play()   }
               else      { bothTo100.play() } });
const condAnimDuration = condAnim.totalDuration() // = 1
\end{js}

\subsection{Relevance of Duration in Other Features}

A wrongly calculated duration becomes more problematic when another feature relies on this calculation. The relative sequencing feature needs the duration of the first animation, so the second animation can be added with the correct offset. For example, we can specify the position parameter \texttt{-=0.5} to specify that it should start 0.5 seonds before the end of the previous animation.

\begin{js}
const bothDelayed = new TimelineMax({ paused: true })
  .add(new TweenMax("#box1", 1, { x: 50 }), 0)
  .add(new TweenMax("#box2", 1, { x: 50 }), "-=0.5");
\end{js}

This feature behaves somewhat unexpectedly when combined with a
conditional animation. In the \texttt{relativeCond} animation below we add a
basic animation followed by a conditional animation. Then we add an animation
with a relative position. The result is that the relative position is
calculated with respect to the duration of the animations before it, which was
a duration of 1.

\begin{js}
const relativeCond = new TimelineMax({ paused: true })
  .add(new TweenMax("#box1", 1, { x: 50 }), 0)
  .add(() => { if (cond) { new TweenMax("#box1", 1, { x: 100 });
               } else { new TweenMax("#box1", 1, { x: 0 }); } })
  .add(new TweenMax("#box2", 1, { x: 50 }), "-=0.5");
\end{js}

Predicting the resulting behavior becomes much more complicated when
conditional animations are embedded deep inside complex timelines and cause
erroneous duration calculations.  Clearly, being more explicit about control
flow structures and their impact on inspectability like in \dsl{} helps
providing more predictable interaction between these features.

\subsection{Relative Sequential Composition in \dsl{}}

While not yet ideal from a usability perspective,\footnote{It
requires
\hs{AllowAmbiguousTypes} (among other extensions) and explicitly instantiating the
constraint \hs{c} at the call-site.
}
 \dsl{} does enable correctly
specifying relative sequential compositions by means of \hs{relSequential}.
\begin{spec}
relSequential :: forall c g.
  (c (Const Duration), c g, Applicative g, Delay g) =>
  (forall f. c f => f ()) -> g () -> Float -> g ()
relSequential anim1 anim2 offset = let
  dur = getDuration (duration anim1)
  in anim1 `sequential` (delay (dur + offset) *> anim2)
\end{spec}

Because this definition requires instances instantiated with \hs{Const Duration},
it only works for animations whose duration can be analyzed. 
Now, we can correctly compose conditional animations sequentially using
relative positioning. We use the \hs{relMaxSequential} function to sequence
animations with a maximum duration.

\begin{spec}
class (LinearTo Float f, IfThenElse f) => Combined f where
instance (LinearTo Float f, IfThenElse f) => Combined f where

relCond :: (LinearTo Float f, IfThenElse f, Applicative f) => f ()
relCond = relMaxSequential @Combined anim1 anim2 (-0.5)
\end{spec}

\section{Future Work}
\label{sec:future}

The current status of \dsl{} is a conceptual design in the space of animation libraries. A logical next step is to put it to the test and aim for the implementation of features provided by a mature animation library such as GSAP.

Another avenue of future work is to explore trade-offs between the mtl-style, as used in this paper, or an initial encoding approach, as is typical in approaches based on algebraic effects and handlers. The mtl-style was chosen since it is simpler presentation-wise, mainly on the extensibility aspect with regards to different computation classes. However, we believe that implementation of some features, such as the relative sequencing, is simpler in the initial approach.

An aspect that was not touched in this paper is \emph{conflict management}. A conflict appears when the same two properties are targeted by different animations in parallel. For example, if we want to change a certain value both to 0 and 100 in parallel, how should this animation look like? \dsl{} does no conflict management, and the animation might look stuttery. GSAP, for example, resolves this by only enabling the most recently added animation, however this strategy is not straightforwardly mapped to the context of \dsl{}.

\section{Related Work}
\label{sec:related}

\paragraph{Functional Reactive Programming}
The origins of functional reactive programming (FRP) lie in the creation of
animations \cite{DBLP:conf/icfp/ElliottH97}, and many later developments use
FRP as the basis for purely functional GUIs. 

\dsl{} focuses on easily describing \emph{micro-animations}, which differ from general
\emph{animoations} as considered by FRP. The latter can typically be described
by a time-paramterized picture function \hs{Time -> Picture}. While a subset of
all possible animations, micro-animations are not easily described by such a
function because many small micro-animations can be active at the same time and
their timing depends on user interaction.

We have only supplied an implementation of \dsl{} on top of a traditional
event-based framework, but it is interesting future work to investigate an
implementation of the \hs{linearTo}, \hs{sequential} and \hs{parallel}
operations in terms of FRP behaviours and events.

\paragraph{Animation Frameworks}

Typical micro-animation libraries for web applications (with CSS or JavaScript)
and animation constructions in game engines provide a variety of configurable
pre-made operations while composing complex animations or integrating new types
of operations is difficult. \dsl{} focuses on the creation of complex sequences
of events while still providing the ability to embed new animation primitives.
We have looked at GSAP as an example of such libraries and some of the limits
in combining extensibility with callbacks and inspectability. \dsl{} is an
exercise in improving this combination of features forward in a direction which
is more predictable for the user.

\paragraph{Planning-Based Animations}

\dsl{} shares similarities with approaches which specify an animation as a plan
which needs to be executed
\cite{DBLP:conf/chi/KurlanderL95,DBLP:conf/eics/MirlacherPB12}. An
animation is specified by a series of steps to be executed, the
plan of the animation. The coordinator, which manages and advances the
animations, is implemented as part of the hosting application. \dsl{} realizes
these plan-based animations with only a few core principles and features the
possibility of adding custom operations and inspection. A detailed comparison
with these approaches is difficult, since their works are very light on
details of the actual implementation aspect.

\paragraph{Inspectable DSLs}

Some DSLs for
parsing~\cite{DBLP:journals/scp/Hughes00,DBLP:journals/corr/CapriottiK14,DBLP:conf/icfp/Lindley14},
non-determinism~\cite{DBLP:journals/corr/abs-1905-06544}, remote
execution~\cite{DBLP:conf/haskell/Gibbons16,DBLP:conf/haskell/GillSDEFGRSS15}
and build systems~\cite{DBLP:journals/pacmpl/MokhovMJ18} focus on inspectability aspects, yet
none of them provide extensibility and expressiveness in addition to inspection.

\section{Conclusion}
\label{sec:conclusion}

We have presented \dsl{}, an extensible and inspectable DSL for micro-animations. \dsl{} focuses heavily on both sequential and parallel composition of animations. This is in contrast with other animation libraries which are mostly focused on sequential composition and only provide a limited form of parallel composition.

We explained the features of \dsl{} with the help of a to-do list application use case. In this use case we showed the use of the additional features of \dsl{}: extensiblity, inspectability and expressivity. We argue that the callback style of providing extensibility hurts the inspectability aspect of animations, which is found in for example the GreenSock Animation Platform.

\end{document}